\documentclass{article} 
\usepackage[french]{babel}
\usepackage[T1]{fontenc}
\usepackage[french]{minitoc}
\usepackage[latin1]{inputenc}
\usepackage{amsmath}
\usepackage{amssymb}
\usepackage[dvips]{color}
\usepackage{graphicx}
\usepackage{subfigure}
\usepackage{makeidx}
\usepackage{setspace}
\usepackage{ulem}
\usepackage{arydshln}
\usepackage{comment}
\DeclareMathSymbol{\R}{\mathbin}{AMSb}{"52}
\DeclareMathSymbol{\N}{\mathbin}{AMSb}{"4E}
\DeclareMathSymbol{\Cx}{\mathbin}{AMSb}{"43}
\DeclareMathSymbol{\Z}{\mathbin}{AMSb}{"5A}

\newcommand{\Et}{\ensuremath{\widetilde{\textbf{E}}}}
\newcommand{\Ht}{\ensuremath{\widetilde{\textbf{H}}}}
\newcommand{\gt}{\ensuremath{\widetilde{g}}} 
\newcommand{\att}{\ensuremath{\widetilde{\alpha}}} 
\newcommand{\htt}{\ensuremath{\widetilde{h}}} 
\newcommand{\etat}{\ensuremath{\widetilde{\eta}}}
\newcommand{\gtb}{\ensuremath{\widetilde{\textbf{g}}}}
\newcommand{\htb}{\ensuremath{\widetilde{\textbf{h}}}}
\newcommand{\Ct}{\ensuremath{\widetilde{C}}}
\newcommand{\ft}{\ensuremath{\widetilde{f}}}
\newcommand{\gamt}{\ensuremath{\widetilde{\gamma}}}

\newcommand{\Mc}{\ensuremath{\mathcal{M}}}

\begin{document}
\doublespacing
\title{Jones matrices of perfectly conducting metallic polarizers}

\author{Philippe Boyer\\\small{D\'epartement d'Optique P.-M. Duffieux, }\\\small{Institut FEMTO-ST, CNRS UMR 6174}
\\\small{Universit\'e de Franche-Comt\'e, 25030 Besan\c con Cedex, France}\\philippe.boyer@univ-fcomte.fr}
\date{}
\maketitle

\begin{abstract}
We deduce from Monomode Modal Method the analytical expressions of transmission and reflexion Jones matrices of an infinitely conducting metallic screen periodically pierced by subwavelength holes. The study is restricted to normal incidence and to the case of neglected evanescent fields (far-field) which covers many common cases. When only one non-degenerate mode propagates in cavities, they take identical forms to those of a polarizer, with Fabry-Perot-like spectral resonant factors depending on bigrating parameters. The isotropic or birefringent properties are then obtained when holes support two orthogonal polarization modes. This basic formalism is finally applied to design compact and efficient metallic half-wave plates.
\end{abstract}

\section{Introduction}
Metallic metamaterials made of subwavelength holes are now designed to exhibit new polarization properties \cite{art:Iwanaga12}. Single periodically pierced metallic screen provides a compact linear polarizer \cite{art:Ren08}, double-layer fishnet metamaterials reveal optical activities \cite{art:Zhang12}, and multi-layer structures allow polarization conversion \cite{art:Xu13}. One topical issue consists in developping efficient theoretical tools to describe with accuracy polarization properties of stacked subwavelength metallic bigrating (SMBG). 

\begin{figure}[htb]
\centerline{\includegraphics[width=10cm]{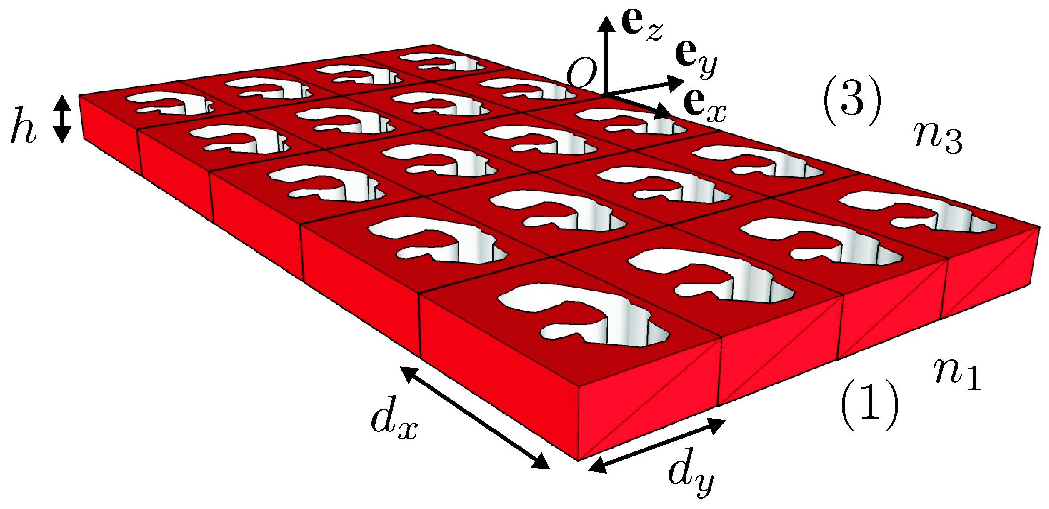}}
\caption{\label{fig:boyerf1} Metallic screen periodically pierced by subwavelength holes.}
\end{figure}

In this paper, we extract from the well-known Momonode Modal Method (MMM) \cite{art:Lalane00,art:Garcia10,art:Boyer12,art:McPhedran77,bk:Petit80} the analytical expressions of reflexion and transmission Jones matrices in far-field approximation of one SMBG pierced by subwavelength holes with arbitrary cross section (see fig. \ref{fig:boyerf1}). The working wavelengths are chosen higher than the first Rayleigh-Wood wavelength in order to consider only one propagative diffracted wave ($0^{th}$-order) in incidence and transmission regions. In accordance with the Jones formalism, we assume that an incident planewave falls on the SMBG in normal incidence. We show that these Jones matrices can be basically expressed in this way:
\begin{equation}
J^{T,R}=\att^{T,R}J_\theta^{(pol,\textbf{e}_x)}-\xi^{T,R}I_d
\label{eq:Jtr}
\end{equation}
where the superscripts $(T,R)$ refer to transmission and reflection respectively, $I_d$ is the identity matrix, $\xi^T=0$ and $\xi^R=1$. The terms $\att^{T,R}$ are Fabry-Perot-like spectral resonant scalar factors and 
\begin{equation}
J_\theta^{(pol,\textbf{e}_x)}=
\left(\begin{array}{cc}
\cos^2\theta & \cos\theta\sin\theta
\\\cos\theta\sin\theta & \sin^2\theta
\end{array}\right)
\end{equation}
is the Jones matrix of a $x$-axis linear polarizer rotated to $\theta$ ($\textbf{e}_x$ being one transverse unit vector of the SMBG and $\theta$ depending on SMBG and pattern rotations, see fig. \ref{fig:boyerf1}). The exact expressions and meanings of $\att^{T,R}$ and $\theta$ will be given in Section 2 (see eq. (\ref{eq:ac})) which deals with the theoretical background of this work. The eq. (\ref{eq:Jtr}) clearly reveals that such a SMBG is equivalent to a spectral resonant linear polarizer in transmission mounting \cite{art:Ulrich63,art:Ulrich67}. Note that this efficient formalism may be used to accurately analyse the important role of the reflected waves via the reflection Jones matrix in multilayer polarizing systems as polarization converters \cite{art:Li10,art:Zhang12,art:Xu13}. The case of rectangular apertures and particularly the role of the pattern rotation in resonance properties are then discussed in Section 3. In Section 4, we extend these equations to study polarization properties of SMBG supporting two modes in holes. We show that the transmission Jones matrix reduces to a sum of two monomode metallic polarizer's Jones matrices when two modes propagates in apertures without mode coupling via evanescent diffracted waves. Consequently, the SMBG behaves as spectral resonant isotropic layer for degenerate modes in one cavity, or reveals birefringence otherwise (two monomode cavities in patterns). Finally, we take advantage of this formalism to design thin half-wave plates with optimized transmission when the SMBG is patterned with two orthogonal rectangular holes \cite{art:Baida11,art:Boutria12}.

\section{Analytical expressions of Jones matrices}

\subsection{Presentation of the problem}
We consider a metallic screen with thickness $h$ periodically pierced by subwavelength holes described in Cartesian coordinate system $Oxyz$ with $(\textbf{e}_x,\textbf{e}_y)$ as unit vectors in transverse plane (determined by bigrating interfaces) and $\textbf{e}_z$ as longitudinal unit vector (see fig. \ref{fig:boyerf1}). We restrict our analyze to biperiodic structures as depicted in fig. \ref{fig:boyerf1} with $Ox$ and $Oy$ as periodic axes, then with $d_x$ and $d_y$ as periods respectively. The metal is assumed perfectly conducting and the refractive index of hole medium is denoted $n_2$. The planar object is surrounded by two semi-infinite homogenenous regions $(j)$ with refractive indices $n_j$, $j\in\left\{1,3\right\}$. An incident plane wave falls on the SMBG from region $(1)$ or $(3)$ in normal incidence and with $\varphi_{inc}$ as polarization incident angle. We introduce the rotation angle $\varphi_{G}$ with the $x$-axis of the SMBG in $Oxy$ plane. The far-field approximation consists to neglect evanescent waves in electromagnetic field description sufficiently distant from the SMBG (half-wavelength about). This hypothesis allows the equivalence between $\varphi_{inc}$ and $\varphi_G=-\varphi_{inc}$ since the light polarization far from the SMBG is given by polarizations of the specular diffracted waves. That is why only the angle $\varphi_G$ is used in the following theory (and not $\varphi_{inc}$) in order to respect the independence of Jones matrices to the incident wave. It is worth noticing that the present theory may easily be used for monoperiodic objects as subwavelength metallic gratings \cite{art:Boyer12} and objects under in oblique incidence.

The present theory is derived from the MMM's basic equations \cite{art:Boyer12} extended to biperiodic structures. The electromagnetic fields are described as Fourier-Rayleigh (FR) expansions in homogenenous regions $(j)$, i.e. as sums of Floquet modes. To simplify notations of FR-orders, $p$-orders stands for $(n,m)$-orders with $n\in[-N,N]$, $m\in[-N,N]$ and $N$ is the truncation order of FR expansions. $p=0$ refers to $(0,0)$-order. Then, we assume that only the non-degenerate fundamental mode ($q=1$) can propagate in cavities (monomode approximation). This hypothesis restricts the spectral validity domain to $\left[\lambda_{c,2},\lambda_{c,1}\right]$ with $\lambda_{c,q}$ the cut-off wavelength of the $q^{th}$ mode. Futhermore, this conditions deals with apertures with $C_i$ ($i\in\N$), $C_{1v}$ or $C_{2v}$ cross section symmetry \cite{art:McIsaac75}. Figure \ref{fig:boyerf2} depicts common examples gathered in $E1$-set as rectangular \cite{art:Mary07,art:Ren08}, ellipsoidal \cite{art:Zhang12}, split-ring \cite{art:Wei11} or chiral \cite{art:Kanda07} hole profiles.

The Jones matrices are now denoted $J_j^T$ and $J_j^R$ when the incident wave is placed in region $(j)$. In the present case of biperiodic objects, the well-known polarizations $(te,tm)$ or $(p,s)$ are used to describe electric field of diffracted FR-waves. Thus, the analytical terms of $J_j^{T,R}$ directly identify to the zero-order (far-field approximation) transmitted and reflected amplitudes given by eqs. (41) and (42) in ref. \cite{art:Boyer12} extended to biperiodic metallic gratings:
\begin{equation}{
J_j^{T,R}=\ft_j^{T,R}
\left(\begin{array}{cc}
\gt_{0,tm}\gt_{0,tm}^* & \gt_{0,tm}\gt_{0,te}^*
\\\gt_{0,te}\gt_{0,tm}^* & \gt_{0,te}\gt_{0,te}^*
\end{array}\right)
-\xi^{T,R}\left(\begin{array}{cc}
1 & 0\\0 & 1
\end{array}\right),
\label{eq:S0}
}\end{equation}
where $\gt_{p,\sigma}$ are the overlap integrals between FR-waves and the cavity modes (expressed below), and $\sigma=\{te,tm\}$ denotes the polarization of the transverse basis-vectors of Floquet mode wavevectors. The factors $\ft_j^{T,R}$ are spectral resonant Airy-like functions:
\begin{equation}
\ft_j^T=\frac{4u\eta_0^{(j)}\etat}{\left[\Ct^{(1)}+\etat\right]\left[\Ct^{(3)}+\etat\right]-u^2\left[\Ct^{(1)}-\etat\right]\left[\Ct^{(3)}-\etat\right]},
\end{equation}
and
\begin{equation}
\ft_j^R=\frac{2\eta_0^{(j)}\left\{\left[\Ct^{(j')}+\etat\right]+u^2\left[\etat-\Ct^{(j')}\right]\right\}}{\left[\Ct^{(1)}+\etat\right]\left[\Ct^{(3)}+\etat\right]-u^2\left[\Ct^{(1)}-\etat\right]\left[\Ct^{(3)}-\etat\right]},
\end{equation}
where
\begin{equation}
\Ct^{(j)}=\sum_p\htb_p^{(j)}\cdot\gtb_p,
\label{eq:Ctj}
\end{equation}
and $j'=1$ if $j=3$ and $j'=3$ if $j=1$. $\eta_{p,\sigma}^{(j)}$ and $\etat$ are the relative admittances of the $p^{th}$ FR-order in region $(j)$ and of the cavity mode respectively. We use the following notations: $\eta_0^{(j)}=\eta_{0,te}^{(j)}=\eta_{0,tm}^{(j)}=n_j$ and $u=exp(i\gamt h)$ where $\gamt$ are the propagation constant of the cavity mode. The terms $\gt_{p,\sigma}$ and $\htt_{p,\sigma}^{(j)}=\eta_{p,\sigma}^{(j)}\gt_{p,\sigma}^*$ for $\sigma\in\{te,tm\}$ are components of vectors $\gtb_p$ and $\htb_p^{(j)}$, respectively. It is worth noticing that coefficients $\Ct^{(j)}$ are computed for $n\in[-N,N]$ and $m\in[-N,N]$, and so it takes into account coupling between cavity mode and evanescent diffracted waves. 

\subsection{Overlap integrals between FR-waves and the cavity modes}
A detailed analysis of $\gt_{p,\sigma}$ expressions provides the basic and analytic formulation of $J_j^{T,R}$ given in eq. (\ref{eq:Jtr}) from the eq. (\ref{eq:S0}). In this aim, we have first to pose transverse (in $(\textbf{e}_x,\textbf{e}_y)$-plane) field expressions of Floquet modes and the ones of cavity modes.

The transverse field profiles of the well-known Floquet modes in homogenenous regions are given by
\begin{equation}
\left\{\begin{array}{l}
\textbf{E}_{p,\sigma}(x,y)=\frac{e^{i\textbf{k}_p.\boldsymbol{\rho}}}{\sqrt{S}}\textbf{e}_{p,\sigma},
\\\textbf{H}_{p,\sigma}(x,y)=\eta_0\eta_{p,\sigma}^{(j)}\textbf{e}_z\wedge\textbf{E}_{p,\sigma}(x,y), 
\label{eq:ExpFR}
\end{array}\right.
\end{equation}
where $\boldsymbol{\rho}=x\textbf{e}_x+y\textbf{e}_y$, $\eta_0=\sqrt{\epsilon_0/\mu_0}$ is vaccum admittance, $S$ the transverse surface area of the periodic cell, $\textbf{k}_p=n2\pi/d_x\textbf{e}_x+m2\pi/d_y\textbf{e}_y$ the transverse component of $p^{th}$ FR-wavevector. The polarization vectors $\textbf{e}_{p,\sigma}$ is
\begin{equation}
\textbf{e}_{p,tm}=
\left\{\begin{array}{l}
\frac{\textbf{k}_p}{\left\|\textbf{k}_p\right\|} \textrm{ if }\left|\textbf{k}_p\right|\neq 0,
\\\cos\varphi_G\textbf{e}_x+\sin\varphi_G\textbf{e}_y \textrm{ if }\left|\textbf{k}_p\right|=0,
\end{array}\right.
\label{eq:eptm}
\end{equation}
and
\begin{equation}
\textbf{e}_{p,te}=
\left\{\begin{array}{l}
\textbf{e}_z\wedge\frac{\textbf{k}_p}{\left\|\textbf{k}_p\right\|} \textrm{ if }\left|\textbf{k}_p\right|\neq 0,
\\-\sin\varphi_G\textbf{e}_x+\cos\varphi_G\textbf{e}_y \textrm{ if }\left|\textbf{k}_p\right|=0.
\end{array}\right.
\label{eq:epte}
\end{equation}
Note that $\left|\textbf{k}_p\right|=0$ is equivalent to $p=0$.

Concerning field expressions inside apertures, we note $\Et(x,y)$ the transverse electric field profiles of the fundamental mode. The corresponding transverse magnetic field $\Ht(x,y)$ is expressed as in eq. (\ref{eq:ExpFR}) substituting $\eta_{p,\sigma}^{(j)}$ by $\etat$ and $\textbf{E}_{p,\sigma}(x,y)$ by $\Et(x,y)$.

Thus, the overlap integrals between FR-orders and the cavity mode are defined by
\begin{equation}
\gt_{p,\sigma}=\iint_{S} \textbf{E}_{p,\sigma}^*(x,y)\cdot\Et(x,y)ds=\textbf{e}_{p,\sigma}\cdot\gtb_p.
\label{eq:gt}
\end{equation}
The integration is only computed on the surface $S$ of cavity cross section since the fields in bigrating are different to zero only on $S$. The $\gtb_p$-vector is
\begin{equation}
\gtb_p=\iint_{S} \Et(x,y)\frac{e^{-i\textbf{k}_p.\boldsymbol{\rho}}}{\sqrt{S}}ds=\gt_p\textbf{v}_p,
\label{eq:gtb}
\end{equation}
$\textbf{v}_p$ is the unit polarization vector of the overlap integrals $\gtb_p$. These overlap vectors cause the linear polarization filtering of the metallic screen which is described in detail below. We introduce the polarization angle $\psi_p$ such as $\textbf{v}_p=\cos(\psi_p)\textbf{e}_x+\sin(\psi_p)\textbf{e}_y$. We easily obtain that 
\begin{equation}
\textbf{e}_{p,te}\cdot\textbf{v}_p=
\left\{\begin{array}{l}
\frac{m d_x\cos(\psi_p)-n d_y\sin(\psi_p)}{\sqrt{n^2d_y^2+m^2d_x^2}} \textrm{ if }\left|\textbf{k}_p\right|\neq 0,
\\-\sin(\varphi_G-\psi_p) \textrm{ if }\left|\textbf{k}_p\right|=0,
\end{array}\right.
\label{eq:projte}
\end{equation}
and
\begin{equation}
\textbf{e}_{p,tm}\cdot\textbf{v}_p=
\left\{\begin{array}{l}
\frac{n d_y\cos(\psi_p)+m d_x\sin(\psi_p)}{\sqrt{n^2d_y^2+m^2d_x^2}} \textrm{ if }\left|\textbf{k}_p\right|\neq 0,
\\\cos(\varphi_G-\psi_p) \textrm{ if }\left|\textbf{k}_p\right|=0,
\end{array}\right.
\label{eq:projtm}
\end{equation}
which is required in eq. (\ref{eq:gt}) to compute $\gt_{p,\sigma}$.

\subsection{Final expressions of Jones matrices}
Making explicit the $0^{th}$ orders overlap integrals finally leads to eq. (\ref{eq:Jtr}) where
\begin{equation}
\att^{T,R}(\lambda,\psi)=\ft_j^{T,R}(\lambda,\psi)\left|\gt_0\right|^2.
\label{eq:ac}
\end{equation}
The $0^{th}$ orders overlap integrals are obtained from eq. (\ref{eq:gt}) in which $\textbf{e}_{p,\sigma}$ are expressed in eqs. (\ref{eq:eptm}) and (\ref{eq:epte}) for $\left|\textbf{k}_p\right|=0$ ($p=0$). The square matrix in eq. (\ref{eq:Jtr}) is the Jones matrix after $\left|\gt_0\right|^2$ being factored out and with $\theta=\psi_0-\varphi_G$. To simplify writing, we introduce the row matrix $\psi=(\cdots,\psi_p,\cdots)$ containing all $\psi_p$-values. The notations highlight here the dependencies of $\ft_j^{T,R}$ and $\att^{T,R}$ on $\lambda$ and $\psi$. Note that coefficients $\att^{T,R}(\lambda,\psi)$ don't depend on $\varphi_G$ since the coupling coefficients $\Ct^{(j)}$ don't (but still vary with $\psi$, see eqs. (\ref{eq:projte}) and (\ref{eq:projtm})). In fact, coefficients $\Ct^{(j)}$ also take the form 
\begin{equation}
\Ct^{(j)}=n_j\left|\gt_0\right|^2+\sum\limits_{p\neq 0}\htb_p^{(j)}\cdot\gtb_p,
\end{equation}
knowing that $\left|\gt_0\right|^2$ and the summation for $p\neq 0$ don't depend on $\varphi_G$.

It is interesting to remark that the determinant of $J_j^T$ given by eq. (\ref{eq:S0}) is equal to zero meaning that the metallic array behaves in transmission as a linear polarizer. This result is confirmed by the final expressions of $J_j^T$ given by eq. (\ref{eq:Jtr}). We can also remark that the resonant factor $\att^T(\lambda,\psi)$ depends on bigrating parameters (via $\textbf{k}_p$-vectors) but polarization properties given by a polarizer's Jones matrix $J_{\psi_0-\varphi_G}^{(pol,\textbf{e}_x)}$ don't. Indeed, the expressions of $J_{\psi_0-\varphi_G}^{(pol,\textbf{e}_x)}$ terms are obtained from eqs. (\ref{eq:projte}) and (\ref{eq:projtm}) for $p=0$ ($\left|\textbf{k}_0\right|=0$). They are thus only depending on $\psi_0$ and $\varphi_G$.

\section{Case of rectangular apertures: Role of pattern rotation}
\begin{figure}[htb]
\centerline{\includegraphics[width=10cm]{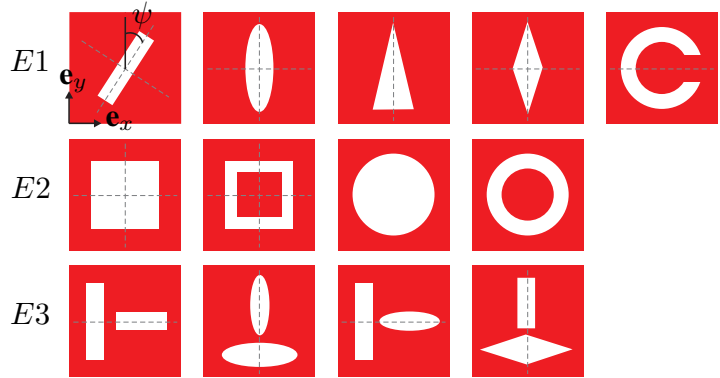}}
\caption{\label{fig:boyerf2} Some examples of common SMBG pattern cross sections considered in the present work. $E1$-set: one hole with one non-degenerate mode. $E2$-set: one hole with two degenerate modes. $E3$-set: two holes, each having a non-degenerate mode.}
\end{figure}

As shown in fig. \ref{fig:boyerf2} ($E1$-set), many hole shapes can be considered. The shape of the apertures first changes the cut-off of each mode and thus causes wavelength shifts of resonance peaks related to cut-off. Secondly, the effective indices of modes (or their admittances particularly required in calculation of overlap integrals) are also affected which equally provokes shifts of Fabry-Perot-like resonance wavelengths. Besides, the radiative losses are changed due to different reflection and transmission conditions at each interface of the bigrating. Consequently, the quality factors are different for each hole shape. We analyse here the influence of the pattern rotation on transmission behaviour of one basic sample. We consider that the SMBG pattern is made of one rectangular hole (first profile of $E1$-set, see fig. \ref{fig:boyerf2}). In this case, analytical expressions of overlap integrals can be easily obtained. For the most other cases (except for the ones for which a pertubtive process may be used), overlap integral calculations directly depending on the transverse field profile require a numerical treatment.

\begin{figure}[htb]
\centerline{\includegraphics[width=10cm]{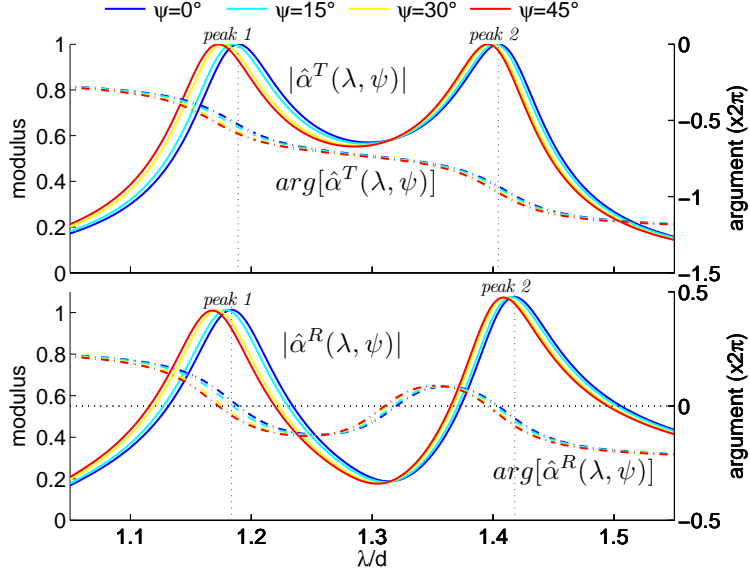}}
\caption{\label{fig:boyerf3} Resonant coefficients $\att^{T,R}(\lambda,\psi)$ versus wavelength for different values of $\psi$. The width of the rectangular hole is $a_x=0.2$ and its length is $a_y=0.7$. Other parameters are $d_x=d_y=1$, $h/d=0.8$, $n_1=n_2=n_3=1$ and $N=5$.}
\end{figure}

We let $d=d_x=d_y$. The width and length are $a_x/d=0.2$ and $a_y/d=0.7$ respectively. $\psi$ identifies here to the angle between the length side and the $\textbf{e}_y$-axis. Other parameters are $h/d=0.8$ and $N=5$. The geometrical parameters are chosen such that resonant peaks appears for $\lambda/d\in[1.0,2.0]$. In this $\lambda$-range, the cavities can be effectively assumed monomode and the polarization angles of $\gtb_p$ is given by the linearly polarized electric field direction of the $TE_{01}$-mode ($\forall p, \psi_p=\psi$). The fig. \ref{fig:boyerf3} depict the spectra of $\att^{T,R}(\lambda,\psi)$ for SMBG patterns rotated by $\psi=0^o$, $15^o$, $30^o$ and $45^o$. The spectra of the terms of the reflexion Jones matrices are directly deduced from those of $\att^R(\lambda,\psi)$ in eq. (\ref{eq:Jtr}). Their analysis reveal interesting properties due to the identity matrix. The fig. \ref{fig:boyerf3} shows two peaks for $\att^R(\lambda,\psi)$ at resonances. This induces common deep peaks (reflexion-like) for diagonal terms of $J_j^R$, whereas the extra-diagonal terms behaves as transmission ones (peaks at resonances). This property may cause some special polarization and transmission effects when several SMBG are piled up. 

\begin{figure}[htb]
\centerline{\includegraphics[width=10cm]{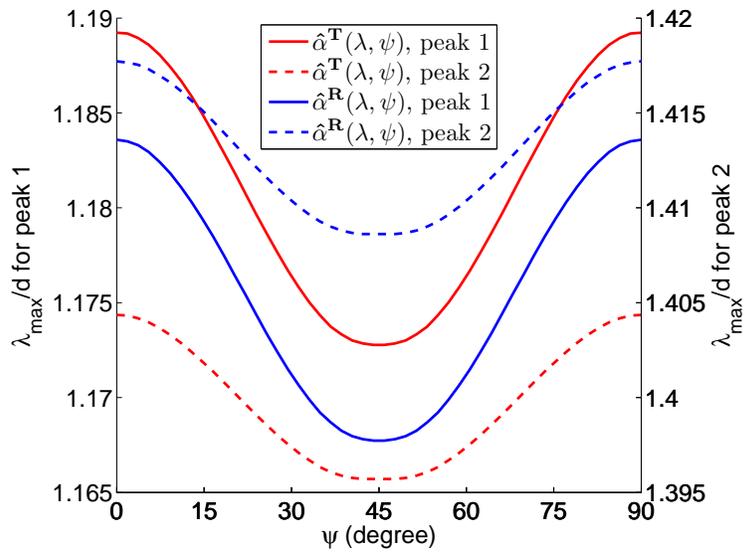}}
\caption{\label{fig:boyerf4} (Color online) Variation of $\lambda_{max}$ at $\left|\att^{T,R}(\lambda,\psi)\right|$ maxima according to $\psi$. See fig. \ref{fig:boyerf3} for parameter values.}
\end{figure}

The peak close to $\lambda/d=1.41$ is related to the resonance at $TE_{01}$-mode cut-off whereas the peak close to $\lambda/d=1.18$ refers to the first Fabry-Perot-like resonance \cite{art:Boutria12}. The resonant functions $\att^{T,R}(\lambda,\psi)$ little depend on SMBG pattern rotation angle $\psi$. The variation of $\psi$ does not affect maxima values but causes wavelength shifts of peaks. The variations of $\lambda_{max}/d$ at each peak maxima of $\att^{T,R}(\lambda,\psi)$ according to $\psi$ are plotted in fig. \ref{fig:boyerf4}. We remark that small blueshifts ($<2.10^{-2}$) occurs when the angle $\psi$ increases from $0^o$ to $45^o$ and redshifts from $45^o$ to $90^o$. The dephasing between diffracted waves and the incident one given by $arg\left[\att^{T,R}(\lambda_{max},\psi)\right]$ (see fig. \ref{fig:boyerf3}) is close to $-\pi$ for first resonant peak ($\lambda/d\approx 1.18$) and $-2\pi$ for the second one ($\lambda/d\approx 1.41$), whereas it remains equal to zero for reflected waves. In view of $\att^{T,R}(\lambda,\psi)$ spectra, we equally show that transmitted and reflected waves don't resonate exactly at the same wavelengths. This may causes special resonance properties of stacks of metallic polarizers.

\section{Extension to bimodal systems}
For some cavity cross sections and/or frequency-ranges, two modes have to be considered in cavities of the SMBG pattern. The first case deals with one cavity allowing two modes (any cross section \emph{a priori}). We highlight the particular case of one degenerate mode (two modes with the same effective index). Knowing that the mode field symmetries are independent of the bigrating lattice's ones only for the studied case of perfectly conducting metals, the degeneracy can be obtained for holes with $C_{iv}$ ($i>2$) cross section symmetry \cite{art:McIsaac75} as circular \cite{art:Ortuno09,art:Nguyen11}, square \cite{art:Mary09,art:Oates13} and annular \cite{art:Baida02,art:Chen13} hole cross sections ($E2$-set in fig. \ref{fig:boyerf2}), or can accidentally occur. The second case deals with two monomode cavities as for any combinaison of geometries in $E1$-set as example. The working spectral range determines the monomode or bimode regime of apertures. However, we focus on the particular case of non-coupled modes via evanescent waves which implies basic expressions of Jones matrices. 

Similarly to eq. (\ref{eq:S0}) for monomode SMBG exposed in Section 2, the eqs.(30), (31), (33) and (34) in ref. \cite{art:Boyer12} applied to bimode holes ($q\in\left\{1,2\right\}$) lead to semi-analytical Jones matrices after tedious calculations:
\begin{equation}
\left\{\begin{array}{l}
J_1^T=2\left[(u\gt)^t M_{11}+\gt^t M_{21}\right]\gt^*\eta^{(1)},
\\J_1^R=2\left[\gt^t M_{11}+(u\gt)^t M_{21}\right]\gt^*\eta^{(1)}-I_d,
\\J_3^T=2\left[(u\gt)^t M_{12}+\gt^t M_{22}\right]\gt^*\eta^{(3)},
\\J_3^R=2\left[\gt^t M_{21}+(u\gt)^t M_{22}\right]\gt^*\eta^{(3)}-I_d,
\end{array}\right.
\label{eq:JTRbi}
\end{equation}
where the superscript $t$ stands for transpose, $u$ is a $2\times 2$-matrix such as $(u)_{q,q'}=u_q\delta_{q,q'}$ with $u_q=exp(i\gamt_q h)$ ($q$ and $q'\in\{1,2\}$), then $(\eta^{(j)})_{\sigma,\sigma'}=\eta_\sigma^{(j)}\delta_{\sigma,\sigma'}$ and $(\gt)_{q,\sigma}=\gt_{q,\sigma}$ with $\gt_{q,\sigma}$ the overlap integral between FR $(0,\sigma)$-order and the $q^{th}$ mode. To simplify notations, we consider here that $\sigma\in\{tm,te\}$ (and $\sigma'$) stands for the subscript $(0,\sigma)$, i.e. for $p=0$ (and $(0,\sigma')$ respectively). The $2\times 2$-matrices $M_{\chi,\chi'}$, with $\chi$ and $\chi'\in\{1,2\}$, are $2\times 2$-blocks of the $4\times 4$-matrix $M=\Mc^{-1}$ with
\begin{equation}
\Mc=\left(\begin{array}{cccc}
\Ct_{1,1}^{(1)}+\etat_1 & \Ct_{1,2}^{(1)} & \left[\Ct_{1,1}^{(1)}-\etat_1\right]u_1 & \Ct_{1,2}^{(1)}u_2
\\\Ct_{2,1}^{(1)} & \Ct_{2,2}^{(1)}+\etat_2 & \Ct_{2,1}^{(1)}u_1 & \left[\Ct_{2,2}^{(1)}-\etat_2\right]u_2
\\\left[\Ct_{1,1}^{(3)}-\etat_1\right]u_1 & \Ct_{1,2}^{(3)}u_2 & \Ct_{1,1}^{(3)}+\etat_1 & \Ct_{1,2}^{(3)}
\\\Ct_{2,1}^{(3)}u_1 & \left[\Ct_{2,2}^{(3)}-\etat_2\right]u_2 & \Ct_{2,1}^{(3)} & \Ct_{2,2}^{(3)}+\etat_2
\end{array}\right).
\label{eq:Mbi}
\end{equation}
This matrix linking field amplitudes of both cavity modes depends on the cross-coupling coefficients $\Ct_{q,q'}^{(j)}$ between $q^{th}$-mode and $q'^{th}$-mode via FR-orders in $(j)$ region. These coefficients are similarly defined as in eq. (\ref{eq:Ctj}):
\begin{equation}
\Ct_{q,q'}^{(j)}=\sum_p\htb_{p,q}^{(j)}\cdot\gtb_{p,q'},
\end{equation}
with $p=(n,m)$, $n\in[-N,N]$ and $m\in[-N,N]$, then $q$ and $q'\in\left\{1,2\right\}$. The numerical inversion of $\Mc$ makes the theory semi-analytical for most cases.

Nevertheless, we are specially interested in the basic form of the Jones matrices in the case of non-coupled modes. Indeed, both modes are not coupled via evanescent waves when their cross-coupling coefficients $\Ct_{1,2}^{(j)}$ and $\Ct_{2,1}^{(j)}$ nullify. Consequently, the matrices $\Mc$ and so $M$ become block-diagonalizable, and $\Mc$ can be analytically inverted. In this case ($\Ct_{1,2}^{(j)}=\Ct_{2,1}^{(j)}=0$), the eqs. (\ref{eq:JTRbi}) and (\ref{eq:Mbi}) lead to the following analytical expressions of Jones matrices:
\begin{equation}
J_j^{T,R}=\sum_{q=1}^2\att^{T,R}(\lambda,\psi_{(q)})J_{\psi_{0,q}-\varphi_G}^{(pol,\textbf{e}_x)}-\xi^{T,R}I_d,
\label{eq:JrtBi}
\end{equation}
with $\psi_{(q)}=(\cdots,\psi_{p,q},\cdots)$. To resume, the transmission Jones matrices of bimode system with non-coupled modes is simply written as the sum of those related to each mode (see eq. (\ref{eq:Jtr})).  

We have now to clarify non-coupling conditions of modes. A tedious analysis of $\Ct_{q,q'}^{(j)}$ terms from eqs. (\ref{eq:projte}) and (\ref{eq:projtm}) shows that two modes are not coupled via evanescent diffracted waves when $\textbf{v}_{p,1}\cdot\textbf{v}_{p,2}=0$ $\forall p$, and when $\textbf{v}_{p,1}$ and $\textbf{v}_{p,2}$ vectors coincide with $\textbf{e}_x$ and $\textbf{e}_y$ respectively. Moreover, the SMBG pattern's cross section must respect $C_{1v}$ symmetry ($E3$-set in fig. \ref{fig:boyerf2}). These assertions reduce to $\psi_{0,1}=0$ and $\psi_{0,2}=\pi/2$ when the mode fields are linearly polarized for which $\psi_{(p)}\equiv\psi$ $\forall p$ (rectangular or square apertures as example). Other geometries inducing $\Ct_{q,q'}^{(j)}=0$ may exist but remain difficult to obtain.  

We thus deduce from the eq. (\ref{eq:JrtBi}) that $J_j^{T,R}$ is a diagonal matrix for $\varphi_G=0$:
\begin{equation}
J_j^{T,R}=\left(\begin{array}{cc}
\att^{R,T}(\lambda,\psi_{(1)})-\xi^{T,R} & 0
\\0 & \att^{R,T}(\lambda,\psi_{(2)})-\xi^{T,R}
\end{array}\right).
\end{equation}
Thus, such metallic plates are divided into two sets:
\begin{enumerate}
\item[i.] $\att^{R,T}(\lambda,\psi_{(1)})=\att^{R,T}(\lambda,\psi_{(2)})$ for the case of one cavity in SMBG pattern with one degenerate mode ($E2$-set). Consequently, the SMBG behaves as an Fabry-Perot-like isotropic resonator in transmission.
\item[ii.] $\att^{R,T}(\lambda,\psi_{(1)})\neq\att^{R,T}(\lambda,\psi_{(2)})$ for other cases, i.e. for patterns made of one cavity allowing two non-degenerate modes ($E1$-set) or two monomode cavities ($E3$-set). Consequently, the SMBG behaves as an Fabry-Perot-like birenfringent resonator in transmission.
\end{enumerate}

\section{Application to metallic half-wave plate}
\begin{figure}[htb]
\centerline{\includegraphics[width=10cm]{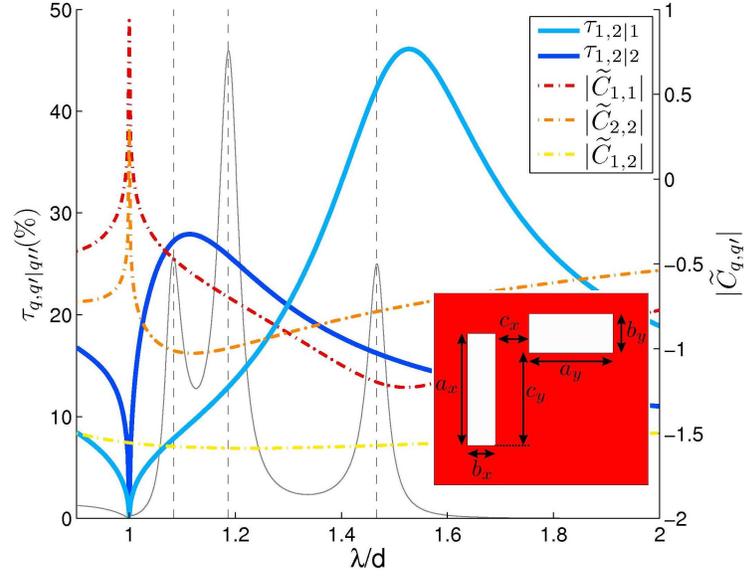}}
\caption{\label{fig:boyerf5} Coupling analysis between modes of each cavities for $L$-shape pattern made of two orthogonal rectangular apertures: $\tau_{q,q'|q''}^{(1)}$ and $\left|\Ct_{q,q'}\right|$ versus $\lambda/d$. The transmission spectrum is plotted in grey color (scale not mentioned).}
\end{figure}

\begin{figure}[htb]
\centerline{\includegraphics[width=10cm]{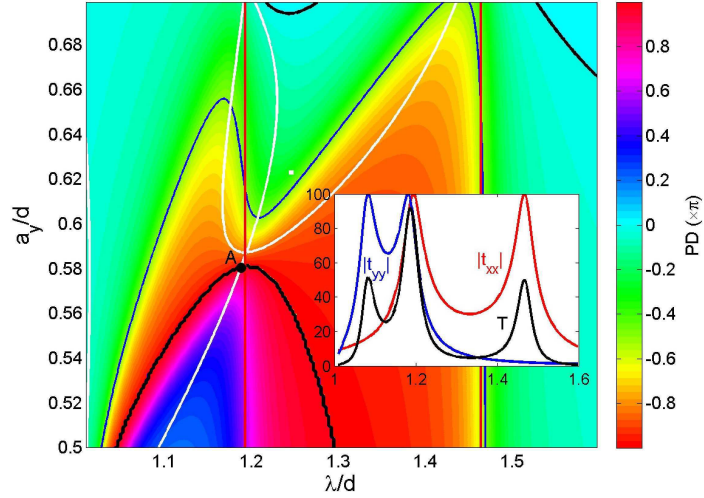}}
\caption{\label{fig:boyerf6} Phase difference PD between $t_{xx}$ and $t_{yy}$ versus $\lambda/d$ and $a_y/d$ for SMBG with $C_{1v}$-pattern made of two orthogonal rectangular apertures. The white contour plots give the couples ($\lambda$,$a_y$) that correspond to $\left|t_{xx}\right|=\left|t_{yy}\right|$. The black line corresponds to $PD=\pi$, the blue line to $PD=\pi/2$ and the red lines to $\left|t_{xx}\right|$ maxima. The point $A$ answers the case of a half-wave plate. The parameter are $h/d=0.83$, $a_x/d=0.73$, $c_x/d=0.067$, $a_y/d=0.58$, $b_y/d=0.2$ and $c_y/d=0.45$.}
\end{figure}

According to the results obtained in the last section, we know that one SMBG with periodic cells made of two orthogonal monomode cavities behaves as an metallic birenfringent plates in transmission which allows design of compact waveplates \cite{art:Baida11,art:Boutria12}. In order to valid our formalism, we consider quasi-identical metallic plates studied in \cite{art:Baida11} but with $C_{1v}$-pattern made of two orthogonal rectangular holes as depicted in fig. \ref{fig:boyerf2} (first pattern of $E3$-set). Actually, the geometry proposed in \cite{art:Baida11,art:Boutria12} including both rectangular profiles does not respect $C_{1v}$-symmetry (but respects $L$-shape) which induces non-nil coupling between modes. We thus introduce the following ratios 
\begin{equation}
\tau^{(j)}_{q,q'|q''}=\left|\Ct_{q,q'}^{(j)}/\Ct_{q'',q''}^{(j)}\right|,
\end{equation}
with $q'\neq q$ to evaluate the significance of these couplings. As shown in Fig. \ref{fig:boyerf5} (calculus made with the complete bimode theory, see eq. (\ref{eq:Mbi})), they are $\tau^{(1)}_{1,2|1}=\tau^{(1)}_{2,1|1}=13\%$ and $\tau^{(1)}_{1,2|2}=\tau^{(1)}_{2,1|2}=25.8\%$ for the case studied in \cite{art:Baida11} at the transmission peak maxima with $\lambda_{max}/d$ close to $1.186$. The values of geometrical parameters are $h/d=0.83$, $a_x/d=0.73$, $c_x/d=0.067$, $a_y/d=0.58$, $b_y/d=0.2$ and $c_y/d=0.45$. All media are filled with air. As in \cite{art:Baida11}, the transmission ($T$) is computed for an incident light with $\varphi_G=45^o$ (and $n_1=n_3$):
\begin{equation}
T=\frac{1}{2}\left(\left|t_{xx}+t_{xy}\right|^2+\left|t_{yx}+t_{yy}\right|^2\right),
\end{equation}
using the same notations introduced in \cite{art:Baida11}. In our theory, $t_{\rho,\rho'}$, $\rho$ and $\rho'\in\left\{x,y\right\}$, are $J_j^T$-terms (given in eq. (\ref{eq:JTRbi})). The structure behaves as a quasi half-wave plate such as the transmission maxima ($T_{max}$) reaches $92\%$ and the phase difference (PD) between $t_{xx}$ and $t_{yy}$ is approximatively equal to $3.15$ radians at $\lambda_{max}$. The transmission Jones matrix is indeed not exactly the same as the one of a perfect half-wave plate knowing that $t_{xy}\neq 0$ and $t_{yx}\neq 0$ since $\Ct_{1,2}^{(j)}\neq 0$ and $\Ct_{2,1}^{(j)}\neq 0$ (see Fig. 3 in \cite{art:Boutria12}). But for our proposed structure, $t_{xy}$ and $t_{yx}$ exactly nullifies when pattern respect $C_{1v}$-symmetry ($c_y=(a_x-b_y)/2$). For this case, the Fig. \ref{fig:boyerf6} depicts the argument of $\att^T(\lambda,0^o)/\att^T(\lambda,90^o)$ versus $\lambda/d$ and $a_y/d$ where $\att^T(\lambda,0^o)$ and $\att^T(\lambda,90^o)$ identify to $t_{xx}$ and $t_{yy}$ respectively (identical parameters as the previous ones). We remark that the figure is similar to Fig. 3 in \cite{art:Baida11}. Point $A$ defines values of $\lambda$ and $a_y$ to achieve one half-wave plate: $arg\left(t_{xx}/t_{yy}\right)\approx\pi$ and $\left|t_{xx}\right|\approx\left|t_{yy}\right|$. The other geometrical parameters have been chosen to maximize transmission maxima (red lines related to $\left|t_{xx}\right|$ maxima).

We propose now to take the advantage of our very efficient analytical model to improve performances of such metallic waveplates. Our goal is to achieve a more compact system (lower thickness) with better transmission. In this aim, the designed object must satisfy to the three following conditions simultaneously:
\begin{equation}
\left\{\begin{array}{l}
L_1=arg(t_{xx}/t_{yy})/\pi-1=0 \textrm{: PD condition,}
\\L_2=\left|t_{xx}\right|-\left|t_{yy}\right|=0 \textrm{: identical transmission moduli condition,}
\\L_3=\left|t_{xx}\right|-1=0 \textrm{: total transmission condition,}
\end{array}\right.
\end{equation}
which are gathered in the global following condition:
\begin{equation}
L=\sum\limits_{l=1}^3\left|L_l\right|.
\end{equation}

\begin{figure}[htb]
\centerline{\includegraphics[width=10cm]{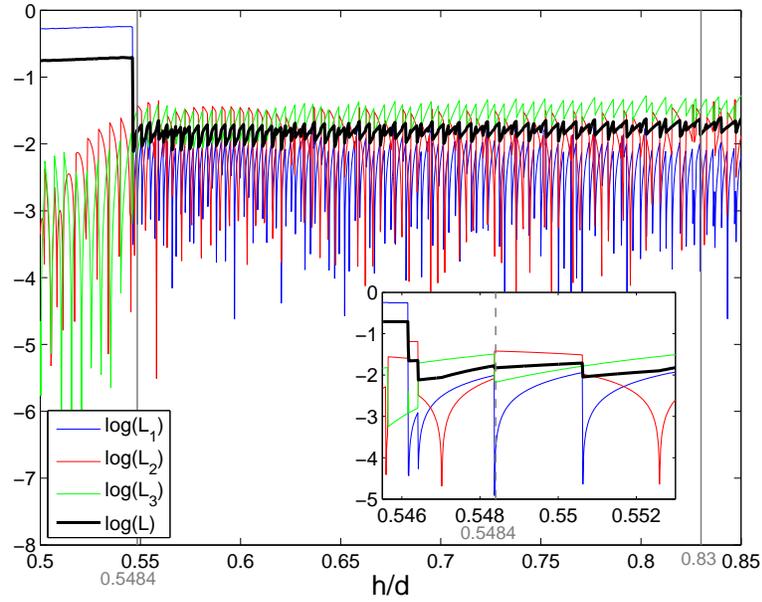}}
\caption{\label{fig:boyerf7} Design of one optimized half-wave plate: $L_1$ to $L_3$ and $L$ computed at point $A$ (see fig. \ref{fig:boyerf6}) as functions of $h/d$. Grey lines refer to equivalent Baida's waveplate ($h/d=0.83$) \cite{art:Baida11} and to the optimized one ($h/d=0.5484$).}
\end{figure}

\begin{figure}[htb]
\centerline{\includegraphics[width=10cm]{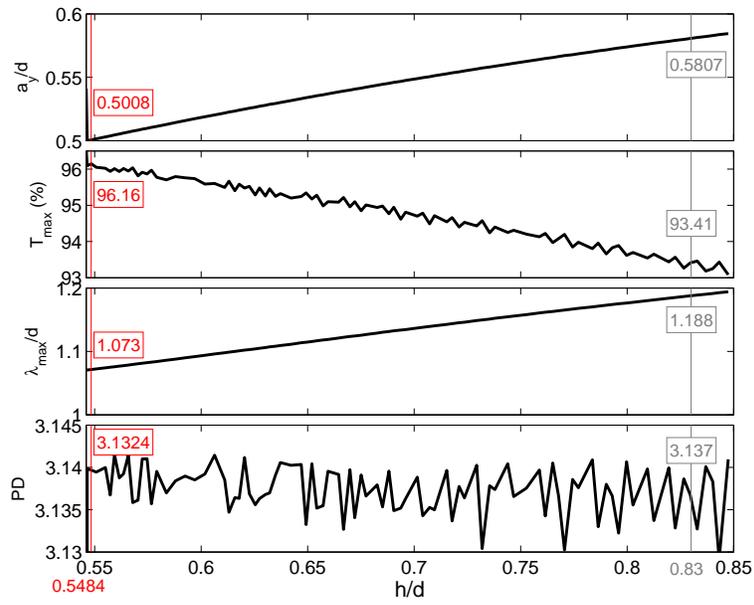}}
\caption{\label{fig:boyerf8} Variations of $a_y/d$, $T_{max}$, $\lambda_{max}/d$ and $PD$ at each minimum of $L$ depicted in Fig. \ref{fig:boyerf7}.}
\end{figure}

\begin{figure}[htb]
\centerline{\includegraphics[width=10cm]{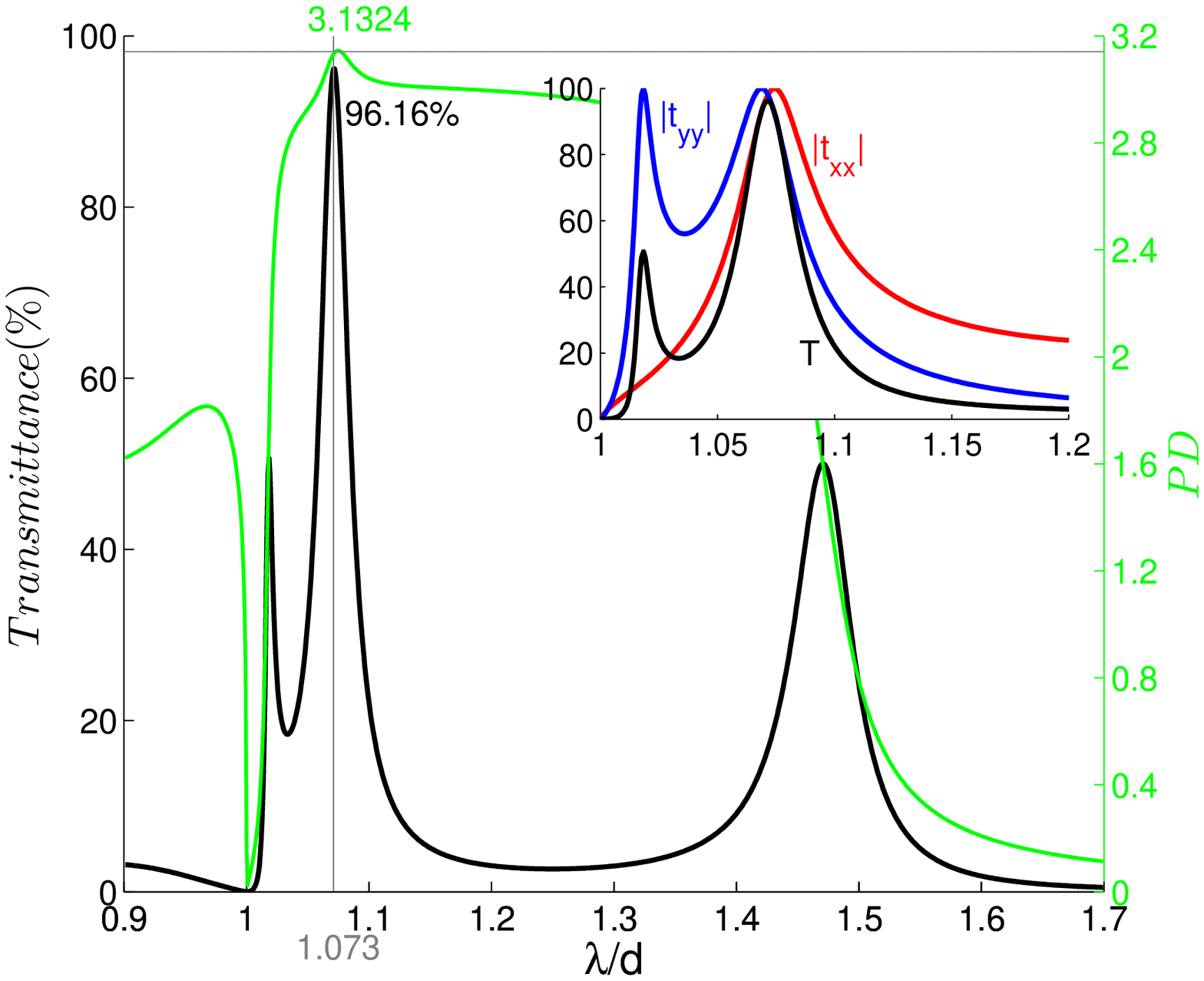}}
\caption{\label{fig:boyerf9} Transmission spectra of the retained metallic plate ($h/d=0.5484$ and $a_y/d=0.5008$).}
\end{figure}

The value of $h/d$ is changed from $0.5$ to $0.85$. For each value of $h$, we determine the point $A$ (and so the values of $\lambda_{max}/d$ and $a_y/d$) as in fig. \ref{fig:boyerf6}. The variations of $L$ and $L_1$ to $L_3$ according to $h/d$ are shown in Fig. \ref{fig:boyerf7}. We see that conditions are satisfied for many values of $h/d$ (hollow peaks). Then, the discontinuities close to $h/d=0.55$ correspond to $a_y/d=0.5$, i.e. when the cut-off wavelength of one cavity mode (position of the first $\left|t_{yy}\right|$ peak) is equal to the Rayleigh wavelengths. Thus, half-wave plate cannot be designed for $h/d<0.55$ about. We also remark that $L_1=0$ and $L_2=0$ cannot occur for the same value of $h/d$ (see subfigure), and $L_3=0$ never occurs. We so deduce that perfect half-wave plates cannot be obtained in general with metallic plates made of subwavelength rectangular holes. The variations of $a_y/d$, $T_{max}$, $\lambda_{max}/d$ and $PD$ at each minimum of $L$ are plotted in Fig. \ref{fig:boyerf8}. In order to achieve our goal, we have chosen the most compact system: $h/d=0.5484$ for one minimum of $L$ such as $L_1\approx 0$ and $L_3$ reaches one of minima. Finally, the transmission of the retained metallic plate is $T_{max}=96.16\%$ at $\lambda_{max}/d=1.073$ and with $PD=3.1324$ rad and $a_y/d=0.5008$. To complete the analysis, the corresponding transmission and $PD$ spectra are plotted in Fig. \ref{fig:boyerf9}. 

\section{Conclusion}
We provide an efficient theoretical tool to analyse polarization features of subwavelength metallic bigratings in monomode and different bimode regimes. The considered geometries cover a wide part of applications studied in litterature. This model has especially been used to optimize thin metallic half-wave plates with high transmission (patterns with two orthogonal rectangular apertures). 
The analytical Jones matrices for one metallic plate and the scattering-matrix propagation algorithm can be combined in an analytical reccurence way. This basic process allows the computation of the global Jones matrices of stacked structures and forms an extended Jones-like formalism for metallic plates. Futher works are in progress to show with the help of this new formalism that an efficient polarization conversion with total transmission occurs for stacked twisted metallic polarizers.

\vspace{1cm}
\emph{Acknowledgements:} I would like to thank Lifeng Li from Dept. of Precision Instruments (Tsinghua University, China) and Daniel Van Labeke from the FEMTO-ST institute (Besan\c con, France) for their helpful advices. 



\clearpage

\clearpage

\clearpage

\clearpage

\clearpage

\clearpage

\clearpage

\clearpage

\clearpage


\begin{thebibliography}{99}

\bibitem{art:Iwanaga12} M. Iwanaga, "Photonic metamaterials: a new class of materials for manipulating light waves", Sci. Technol. Adv. Mater. \textbf{13,} 053002 (2012).

\bibitem{art:Ren08} X.-F. Ren, P. Zhang, G.-P. Guo, Y.-F. Huang, Z.-W. Wang and G.-C. Guo, "Polarization properties of subwavelength hole arrays consisting of rectangular holes", Appl. Phys. B \textbf{91,} 601--604 (2008).

\bibitem{art:Zhang12} Y.-L. Zhang, W. Jin, X.-Z. Dong, Z.-S. Zhao and X.-M. Duan, "Asymmetric fishnet metamaterials with strong optical activity", Opt. Express \textbf{20,} 10776--10787 (2012).

\bibitem{art:Xu13} J. Xu, T. Li, F.F. Lu, S.M. Wang and S.N. Zhu, "Manipulating optical polarization by stereo plasmonic structure", Opt. Express \textbf{19,} 748--756 (2011).

\bibitem{art:Lalane00} P.~Lalanne, J.P. Hugonin, S.~Astilean, M.~Palamaru, and K.D. Möller, "One-mode model and Airy-like formulae for one-dimensional metallic gratings", J. Opt. A: Pure Appl. Opt. \textbf{2,} 48--51 (2000).

\bibitem{art:Garcia10} F.J. Garcia-Vidal, L. Martin-Moreno, T. W. Ebbesen and L. Kuipers, "Light passing through subwavelength apertures", Rev. Mod. Phys. \textbf{82,} 729--787 (2010).

\bibitem{art:Boyer12} P. Boyer and D. Van Labeke, "Analytical study of resonance conditions in planar resonators", J. Opt. Soc. Am. A \textbf{29,} 1659--1666 (2012).

\bibitem{art:McPhedran77} R.C. McPhedran,D. Maystre, "On the theory and solar application of inductive grids", Appl. Phys. \textbf{14,} 1--20 (1977).

\bibitem{bk:Petit80} R. Petit, "Electromagnetic Theory of Gratings, Topics in Current Physics", \textbf{22,}, Springer-Verlag, Berlin, 1980.

\bibitem{art:Ulrich63} R. Ulrich, K.F. Renk, L. Genzel, "Tunable Submillimeter Interferometers of the Fabry-Perot Type", Microwave Theory and Techniques, IEEE Transactions on., \textbf{11,} 363--371, 1963.

\bibitem{art:Ulrich67} R. Ulrich, "Far-infrared properties of metallic mesh and its complementary structure", Infrared Physics. \textbf{7,} 37--55 (1967).

\bibitem{art:Li10} T. Li, S.M. Wang, J.X. Cao, H. Liu and S.N. Zhu, "Cavity-involved plasmonic metamaterial for optical polarization conversion", App. Phys. Lett. \textbf{97,} 261113 (2010).

\bibitem{art:Baida11} F.I. Baida, M. Boutria, R. Oussaid and D. Van Labeke, "Enhanced-transmission metamaterials as anisotropic plates", Phys. Rev. B \textbf{84,} 035107 (2011).

\bibitem{art:Boutria12} M. Boutria, R. Oussaid, D. Van Labeke and F.I. Baida, "Tunable artificial chirality with extraordinary transmission metamaterials", Phys. Rev. B \textbf{86,} 155428 (2012).

\bibitem{art:McIsaac75} P.R. McIsaac, "Symmetry-induced modal characteristics of uniform waveguides I: summary of results", Microwave Theory and Techniques, IEEE Transactions on  \textbf{23,} 421--429 (1975).

\bibitem{art:Mary07} A. Mary, S.G. Rodrigo, L. Martin-Moreno, and F.J. García-Vidal., "Theory of light transmission through an array of rectangular holes", Phys. Rev. B \textbf{76,} 195414 (2007).

\bibitem{art:Wei11} Z. Wei, Y. Cao, Y. Fan, X. Yu and H. Li, "Broadband polarization transformation via enhanced asymmetric transmission through arrays of twisted complementary split-ring resonators", Appl. Phys. Lett. \textbf{99,} 221907 (2011).

\bibitem{art:Kanda07} N. Kanda, K. Konishi and M. Kuwata-Gonokami, "Terahertz wave polarization rotation with double layered metal grating of complimentary chiral patterns", Opt. Express \textbf{15,} 11117 (2007).


\bibitem{art:Ortuno09} R. Ortuno, C. Garcia-Meca, F.J. Rodriguez-Fortuno, J. Marti and Alejandro Martinez, "Role of surface plasmon polaritons on optical transmission through double layer metallic hole arrays", Phys. Rev. B \textbf{79,} 075425 (2009).

\bibitem{art:Nguyen11} T.D. Nguyen, S. Liu, Z.V. Vardeny and A. Nahata, "Engineering the properties of terahertz filters using multilayer aperture arrays", Opt. Express \textbf{19,} 18678-18686 (2011).

\bibitem{art:Mary09} A. Mary, S.G. Rodrigo, L. Martín-Moreno and F.J. García-Vidal, "Holey metal films: From extraordinary transmission to negative-index behavior", Phys. Rev. B \textbf{80,} 165431 (2009).

\bibitem{art:Oates13} T.W.H. Oates, B. Dastmalchi, C. Helgert, L. Reissmann, U. Huebner, E.-B. Kley, M.A. Verschuuren, I. Bergmair, T. Pertsch, K. Hinger and K. Hinrichs, "Optical activity in sub-wavelength metallic grids and fishnet metamaterials in the conical mount", Opt. Mat. Express \textbf{3,} 439--451 (2013).

\bibitem{art:Baida02} F.I. Baida and D. Van Labeke, "Light transmission by subwavelength annular aperture arrays in metallic films", Opt. Comm. \textbf{209,} 17--22 (2002).

\bibitem{art:Chen13} Z. Chen,C. Wang, Y. Lou,B. Cao and X. Li, "Quarter-wave plate with subwavelength rectangular annular arrays", Opt. Comm. \textbf{297,} 198--203 (2013).

\end{thebibliography}
\end{document}